# Functional Clustering of Discount Functions for Behavioural Investor Profiling

Annamaria Porreca[1], Viviana Ventre[2], Roberta Martino[2],

Salvador Cruz Rambaud[3], Fabrizio Maturo[1*]

[1]Department of *Economics, Statistics and Business*, Universitas *Mercatorum*, Rome, Italy.
[2]Department of *Mathematics and Physics*, University of Campania *Luigi Vanvitelli*, Caserta, Italy.
[3]Department of *Economics and Business*, University of Almeria, Almeria, Spain.

**ABSTRACT**

Classical finance models are based on the premise that investors act rationally and utilize all available information when making portfolio decisions. However, these models often fail to capture the anomalies observed in intertemporal choices and decision-making under uncertainty, particularly when accounting for individual differences in preferences and consumption patterns. Such limitations hinder traditional finance theory's ability to address key questions like: How do personal preferences shape investment choices? What drives investor behaviour? And how do individuals select their portfolios? One prominent contribution is Pompian's model of four Behavioral Investor Types (BITs), which links behavioural finance studies with Keirsey's temperament theory, highlighting the role of personality in financial decision-making. Yet, traditional parametric models struggle to capture how these distinct temperaments influence intertemporal decisions, such as how individuals evaluate trade-offs between present and future outcomes. To address this gap, the present study employs Functional Data Analysis (FDA) to specifically investigate temporal discounting behaviours revealing nuanced patterns in how different temperaments perceive and manage uncertainty over time. Our findings show heterogeneity within each temperament, suggesting that investor profiles are far more diverse than previously thought. This refined classification provides deeper insights into the role of temperament in shaping intertemporal financial decisions, offering practical implications for financial advisors to better tailor strategies to individual risk preferences and decision-making styles.

---

* Corresponding author: fabrizio.maturo@unimercatorum.it

## 1. INTRODUCTION

The focus of financial studies is increasingly shifting towards analysing the behaviours of agents within financial systems (e.g., investors and traders) rather than solely examining the characteristics of the markets themselves (such as bonds, forex, or stock markets). The emergence of behavioural finance as a new field of investigation is motivated by the interest and the need to understand how the psychobiological characteristics of the investor can influence the way financial decisions are made (Joo and Kokab, 2015). In this regard, traditional finance theory plays a limited role because it assumes that the investor has no problems understanding information and managing emotional factors in the portfolio investment decision-making process. The integration of cognitive psychology into the study of decision-making dynamics has defined the foundations of behavioural finance: the condition of bounded rationality (Simon, 1990) and the presence of heuristics and behavioural biases (Kahneman and Tversky, 1979; Tversky and Kahneman, 1989). Since the individual's personality influences decision-making, behavioural finance has turned to various personality theories to determine the investor's characteristics (Rao and Lakkol, 2022). An application of personality theories to financial choices is provided by the work of Pompian (2012, 2016), who defined a model of four types of behavioural investors (BITs), combining the studies of behavioural finance with Keirsey's model (Levorin, 2021). Keirsey's temperament model (Keirsey and Bates, 1984) is based on defining four classes of individuals: Artisan, Guardian, Idealist and Rational. The four temperaments are based on Jung's neo-psychoanalytic theory, and the four variants are related to the sixteen Myers-Briggs types (Myers et al., 1998; Rao and Lakkol, 2022; Robbins and Ross, 2020). The Keirsey Temperamental Model (KTM) debuted in behavioural finance with the study by Wheeler et al. (2002), in which the authors estimated the impact of personality on the performance of accounting professionals. The second study, which sought to provide conceptual and empirical evidence of the link between KTM and decision-making, was conducted by McKenna et al. (2003). The authors concluded that individual temperament influences financial decisions. Subsequently, not only did Statman and Wood (2004) identify a relationship between risk attitude and temperament, but Pan and Statman (2010) also investigated how temperaments can influence biases. Furthermore, Parsaeemehr et al. (2013) even found that temperament can affect the reception of financial information. In the behavioural context, and that is why we decided to use it, the Kersey Temperament Sorter II (Keirsey, 1998) distinguishes itself by focusing on behaviour rather than preferences (Robbins and Ross, 2020) and is helpful for financial advisors in analysing clients (Statman and Wood, 2004). A fundamental line of research in behavioural finance for client classification is risk management, which refers to risk capacity and risk appetite: the former indicates the risk an individual can tolerate; the latter suggests the individual's ability to take on risk (Pompian, 2016, 2017). In this regard, Pompian (2016, 2017) points out that in a behavioural context, it is crucial to discuss risk appetite and risk capacity in terms of known and unknown risks: "When clients can at least understand and measure risks they are taking (i.e., known risks), they can accept the results. When the risks they believe they accepted include outcomes that are outside the bounds of what they expect or can reasonably understand (i.e., unknown risks), behavioural problems often begin." (Pompian, 2016, p. 4-5). Uncertainty management is fundamental to investigating how individuals interface with investment decisions (Geng et al., 2022) to mitigate its effects. Understanding how perceptions of the future's indeterminacy impact the outcomes assessment is crucial. Intertemporal choice theory, whose mathematical reference model is the Discounted Utility Model (DUM) (Samuelson, 1937, 1952), provides a quantitative and qualitative approach for describing intertemporal discounting, i.e., when decisions involving trade-offs among costs and benefits occur at different times (Frederick et al., 2002). The characteristics of intertemporal discounting are related to individual aspects such as psychological, socio-demographic, and genetic variables (Keidel et al., 2021). Even in intertemporal choices, the limited rationality condition, heuristics, and behavioural biases make individuals less rational than they think.

Anomalies in the DUM (Loewenstein and Prelec, 1992), i.e., those preferences that deviate from the normative model of exponential intertemporal discounting, have been investigated through the psychological mechanisms responsible for decision-making (Prelec, 2004; Takahashi et al., 2008; Ventre and Martino, 2022; Ventre et al., 2022a Ventre et al., 2022b). In this context, a discount hyperbolic function is closer to the actual behaviour of individuals (Green and Myerson, 1997). Ventre et al. (2023) investigated the relationship between Keirsey's four temperaments and the degree of decrease in impatience (Prelec, 2004) through the hyperbolic factor (Rohde, 2010). The latter expresses the emotional drives and cognitive distortions involved in evaluating alternatives.

The present paper aims to improve and deepen the link between the characteristics of the discount function and Keirsey's four temperaments by using Functional Data Analysis (FDA) combined with functional clustering techniques as investigative tools. There are two main reasons for this methodological combination. First, FDA proves to be an innovative tool for investigating the characteristics of intertemporal preferences, as it overcomes the limitations defined by traditional parametric approaches. FDA enables a comprehensive interpretation of the data by capturing both the mathematical properties of the discount function and the interrelationships between the factors that influence the decision-making process. Secondly, the integration of functional clustering extends the analysis by revealing hidden patterns and subgroups within the main behavioural investor types (BITs). This approach allows us to go beyond a mere descriptive analysis of discrete preference points and identify nuanced temporal discounting behaviours that may not be apparent through standard methods.

The application of FDA was made possible by introducing the uncertainty aversion measure, defined and expressed empirically by Ventre and Martino (2022). To showcase the potential of the FDA approach concerning intertemporal choices in behavioural finance, we collected 200 interviews via an online questionnaire. The data analysis combines the descriptive examination of intertemporal choice theory, based on decreased impatience and uncertainty aversion, with the results from FDA and functional clustering. Our findings confirm that the four temperaments exhibit distinct discounting behaviours and that high heterogeneity exists within each temperament group, thereby validating the importance of identifying subpatterns in behavioural profiling under uncertainty.

The paper is structured as follows: Section 2 introduces the Keirsey temperament model and behavioural investor types; Sections 3 and 4 present the method of analysis, including intertemporal choice theory, FDA, and clustering techniques; Section 5 presents the experimental procedure and results, followed by the discussion and conclusion in which possible applications and future research directions are proposed.

## 2. BACKGROUND ON THE KEIRSEY TEMPERAMENT MODEL (KTM) AND BEHAVIORAL TYPES OF INVESTORS

Behavioural finance has turned to different approaches to personality psychology (Rao and Lakkol, 2022). The theory in the present paper is Keirsey's Temperaments (Keirsey and Bates, 1984), whereby a score of four different temperaments characterises the individual. Briefly (Statman and Wood, 2004; Robbins and Ross, 2020; Yilmaz and O'Connor, 2015): Guardians have a natural talent for management, used to make things work well in social and public life. They follow rules and cooperate with others with prudence, loyalty, and discipline. Artisans emerge in the arts, from artistic to business. They tend to be the centre of attention, are generally impulsive and competitive, do not prefer to wait, love freedom, and live to "carpe diem" (Yilmaz and O'Connor, 2015, p. 130). Idealists are concerned with their personal growth, aiming to become the best version of themselves. They are inclined to work with people in the educational and social fields. Individuals of this temperament are known to be the most communicative type. Rationales are analytical and logical, like to solve complex problems, apply logic and rationality to their activities, and are constantly motivated to understand how things work. This temperament is known for its logical skills and ability to solve problems quickly, and it is characterised by the tendency to be sceptical, pragmatic, and independent.

Pompian (2012), cross-referencing behavioural finance studies with Keirsey's temperaments, defined four behavioural investor types (BITs) (Cervellati, 2018). The relationship between Keirsey

Temperament Theory, Behavioral Investor Types and Myers-Briggs types is presented in Table 1 (Rao and Lakkol, 2022; Ventre et al., 2023; Cervellati, 2017).

| Keirsey Temperament Theory | Behavioral Investor Types | MBTI |
|---|---|---|
| Artisan | Accumulator | SP |
| Guardian | Preserver | SJ |
| Idealist | Follower | NF |
| Rational | Independent | NT |

**Table 1.** Relationship between KTM and BITs.

A Preserver is a guardian of their assets and suffers significantly from losses. They hardly act out of fear of making the wrong decision or risking too much. Therefore, they prefer the status quo and avoid making decisions. This type of investor focuses on the short term, and the main concern is not to lose what has been gained previously. From a practical perspective, Preservers are challenging to advise, and because they react emotionally to fluctuations in the value of their portfolios, they avoid losses. Preservers need much support and could benefit from behavioural coaching rather than a rigid education in finance and investment. It is essential to understand how the portfolio responds to emotional needs, such as those of the family. By developing a trusting relationship, Preservers become loyal clients by appreciating the advisor's expertise. The Follower investor is passive, follows trends and friends, and often lacks a long-term plan. They may react differently to proposed investments depending on how they are presented, and they usually increase the risk of loss by following trends without assessing valuations. Follower investors often overestimate their risk appetite, following trends and making investments that may not be ideal in the long term. Advisers must encourage them to carefully evaluate behavioural trends and educate them about diversifying and planning for the long term. Loyalty and adherence to long-term investment plans are fostered by challenging them to think through and justify decisions based on data. Independent investors are autonomous. They have original ideas and actively participate in the financial markets. They are critical and willing to take risks. Their decisions are usually based on analysis and logic. However, they can be influenced by behavioural biases such as cognitive conservatism and the tendency to seek confirmation of their opinions. They are financially literate and tend to do extensive research before investing. However, they may find it difficult to admit mistakes. Independents, although challenging to advise because of their autonomy, will accept advice that respects their opinions. In this case, education is vital to change their cognitive biases. Polite and clear discussions during meetings can lead to positive changes. An accumulator is an investor determined to grow his wealth; they tend to have a risk appetite, a tendency to take the lead in decision-making, and perhaps overconfidence that they can control what they invest in. On the other hand, overconfidence can result in impulsiveness, a high appetite for risk, and profound unease when things go wrong. Basic principles such as diversification and building solid relationships with financial advisers may be complex for these investors. Accumulators are emotionally involved in investment decisions and may reject advice that limits their risk tolerance. Advisers should be their guide, and cooperation is encouraged by demonstrating the ability to support informed long-term decisions.

The characteristics discussed can be summarised in Table 3 (Pompian, 2012, p.103, 111,121, 135).

| BIT | Basic Orientation | Dominant Bias Types | Impactful Biases | Investing Style | Level of Risk Tolerance |
|---|---|---|---|---|---|
| **Preserver** | Loss-averse and deliberate in decision-making | Emotional, relating to fear of losses and inability to make decisions and act | Loss Aversion and Status Quo | Wealth preservation and Status Quo | Generally lower than average |
| **Follower** | General lack of interest in money and investigating and typically desires direction when making financial decisions | Cognitive, relating to following behaviour | Recency and Framing | Passive | Generally lower than average but often thinks risk tolerance level is higher than it is |
| **Independent** | Engaged in the investment process and opinionated on investment decisions | Cognitive, relating to some pitfalls associated with doing one's own research | Confirmation and Availility | Active | Generally, above but not so high as aggressive investors |
| **Accumulator** | Interested and engaged in wealth accumulation and confident in investing ability | Emotional, relating to overconfidence and desire for influence in the investment process | Overconfidence and illusion of control | Actively engaged in decision-making | High to very high |

**Table 2**. Characteristics of BITs described by Pompian (2012).

## 3. TREATING INTERTEMPORAL DECISION-MAKING PROFILES VIA FUNCTIONAL DATA ANALYSIS

### 3.1 Construction of the individual discount function

The Discounted Utility Model is essential for studying decision-making behaviour in intertemporal choices. It predicts that given an intertemporal prospect $(x_1, t_1; \ldots; x_n, t_n)$ composed by $n$ alternatives $x_i$ available at time $t_i$, intertemporal utility is defined as $U(x_1, t_1; \ldots; x_n, t_n) =$ $\sum_i^n U(x_i)f(t_i)$ in which $U(x_i)$ is the cardinal utility of $x_i$ and $f(t_i)$ is the discount function evaluated at the points where the decision-maker will receive the outcome. In other words, the discount function reduces the present utility of a good according to how the individual perceives future indeterminacy. Thus, the decision maker will prefer the prospect $(x_1, t_1; \ldots; x_n, t_n)$ to $(y_1, s_1; \ldots; y_m, s_m)$ if $\sum_i^n U(x_i)f(t_i) > \sum_i^m U(y_i)f(s_i)$ and will be indifferent if $\sum_i^n U(x_i)f(t_i) = \sum_i^m U(y_i)f(s_i)$.

The behavioural aspects of the discount functions are caught by the value of the discount rate (Read, 2004) and the degree of impatience (Cruz Rambaud and Muñoz Torrecillas, 2016), which are assumed constant in Samuelson's normative model. In contrast, the hyperbolic model is characterised by a steeper discount in periods close to the present, and is quantitatively identified by a degree of impatience and a discount factor that decreases over time. These characteristics were investigated through a time transformation that shifts the gap between expected and empirical preferences from the value of the discount function to the subjective perception of time (Ventre and Martino, 2022; Ventre et al., 2022b). In this way, it is possible to investigate the characteristics of the discount function through the interpolation of $n$ values obtained after administering a questionnaire, leading to the time course of the curve for each individual. This construction determines a measure which quantifies the decision maker's uncertainty aversion (Ventre and Martino, 2022). Table 3 presents the questionnaire based on the discussed approach and used for the empirical data collection (Martino et al., 2024; Ventre et al. 2024).

In addition, to further investigate the relationship between intertemporal preferences and Keirsey temperaments, the values of individual hyperbolic factors will also be investigated, as discussed in Ventre et al. (2023). In particular, considering a pair of indifference $(x, s) \sim (y, t)$ and $(x, s + \sigma) \sim (y, t + \tau)$, with $s < t$, $x < y$, $\sigma > 0$ and $\tau > 0$, the hyperbolic factor is defined as:

$$H(s, t, \sigma, \tau) := \frac{\tau - \sigma}{t\sigma - s\tau} \qquad [1]$$

which is a measure of the degree of decreasing impatience (Rohde, 2010). The reason for considering this measure in the present work lies in the research conducted by Ventre et al. (2022a) in which the mechanism of decreasing impatience was related to the description of some anomalies in the Discounted Utility Model and discussed concerning possible behavioural biases involved.

| | Question | Function |
|---|---|---|
| $U_D(x(0))$ fixed $t_0 = 0$ $t = 2,4,7,10,14,$ $20,30,45,60,90$ | You should receive $U(x(t_i))$ euros today, how much do you want to receive in $t_{i+1}$ days to consider the offer equivalent? | $f(t) = \begin{cases} f(0) = 1 \\ f(t_{i+1}) = \dfrac{f(t_i) \cdot U(x(t_i))}{U(x(t_{i+1}))} \end{cases}$ |

**Table 3**. Questionnaire used for interpolation of the individual discount functions (Martino et al., 2024; Ventre et al. 2024).

### 3.2 FDA in intertemporal preferences

To fit the available data to a wide range of discount functions, reflecting different degrees of inconsistency, we propose Functional Data Analysis (FDA). Each vector of pairs of indifferences corresponding to the responses of individuals can be viewed as a function in the time domain rather than as a vector in a finite dimension. The sequences of individual discrete observations can be represented as functions and analysed as a single entity. If sample paths belong to a finite-dimension space spanned by a basis, the function $f(t)$ could be represented by a basis expansion:

$$f(t) = \sum_{k=1}^{K} a_{ik}\phi_k(t) \quad i = 1, \cdots, n \qquad [2]$$

where $f_i(t)$ is the reconstructed function for the $i$-th unit (time points could be different for each unit); $\phi_k(t)$ are linearly independent and known functions, called basis functions; and finally, $a_{ik}$ are coefficients that link each basis function together in the representation of $f_i(t)$. However, to be considered a discount function, a b-spline approximation must also satisfy the following conditions: $f(0) = 1$; $f(t) > 0$; and $f(t)$ is strictly decreasing. Hence, we turn to the method of monotone smoothing introduced by Ramsay and Silverman in 2005. When functions must meet specific constraints such as positivity and monotonicity, it becomes challenging to confine linear combinations of basis functions for this purpose. Ramsay proposed a solution by transforming the issue into one where the estimated curves are not restricted (Ramsay, 2005). In certain situations, there is a need for a fitting function *f(t)* that either increases or decreases monotonically, even if the observed data might not display perfect monotonic behaviors:

$$y_j = b_0 + b_1 f(t_j) + e_j. \qquad [3]$$

To solve this problem, we can express $Dx$ as the exponential of an unconstrained function $W$ to obtain:

$$Df(t) = \exp[w(t)]dt. \qquad [4]$$

By integrating both sides of this equation, we get:

$$f(t) = \int_{t_0}^{t} \exp[w(u)]du, \qquad [5]$$

where $t_0$ is the fixed origin for the range of $t$-values for which the data are being fit. The intercept term $b_0$ is the value of the approximating function at $t_0$ (Ramsay, 2005) (in this context it is $f(0) =$ 1). To accommodate monotonically decreasing functions, Ramsay (2005) proposed the idea of keeping $b_1$ distinct and opting to normalise $w(u)$ for numerical stability. Upon substitution, we obtain:

$$y_j = b_0 + b_1 \int_{t_0}^{t_j} \exp[w(u)]du + e_j \qquad [6]$$

The function $w(u)$ is now the logarithm of the data-fitting function $f(u) = \exp[w(u)]$.

## 4. FUNCTIONAL CLUSTERING IN THE CONTEXT OF INTERTEMPORAL DECISION-MAKING

Functional clustering is a powerful method designed to handle complex data structures where observations are represented as curves, trajectories, or other functional forms (Ramsay, 2005). Unlike traditional clustering methods, which are suitable for point-wise data, functional clustering aims to group entire functions, capturing the temporal or continuous nature of the underlying phenomena.

In this section, we expand the concept of functional clustering to analyse investor behaviour, viewing each investor's decision-making pattern as a continuous function over a specific period. By applying functional clustering techniques, we aim to uncover hidden patterns within investor groups, offering a nuanced understanding of decision-making behaviours. This methodology enables us to investigate subtle differences between investor types, using a lens that captures the evolution and dynamics of behaviours in a continuous space rather than a static, point-wise comparison. By treating investor preferences as functional data, we shift the analysis towards a richer representation that allows us to explore heterogeneity in decision-making with a more refined granularity, ultimately contributing to a deeper understanding of behavioural finance.

### 4.1 Proximity Measures and Functional Mean and Variability in the Context of Investor Analysis

Proximity measures can be pivotal in analysing individual investor behaviour in unsupervised classification (clustering). The choice of a specific proximity measure depends on the nature of the functional data and the study's particular objectives. Different metrics and semi-metrics can be applied to perform clustering on investor profiles in the context of FDA (Ramsay, 2005). Focusing on the $L_2$-space, the most used distances for functional data are:

1. The $L_2$-distance:

$$|f_1(t) - f_2(t)|_2 = \left\{ \frac{1}{\int_a^b w(t)\,dt} \int_a^b |f_1(t) - f_2(t)|^2 w(t)\,dt \right\}^{\frac{1}{2}} \qquad [7]$$

where *w(t)* is a strictly positive weight function, and the points on each curve are equally spaced. This distance is used when focusing on the overall difference between all the curves.

2. The semi-metric of the *r*-order derivatives of two curves $f_1(t)$ and $f_2(t)$:

$$d_2^{(r)}(f_1(t), f_2(t)) = \left[1/T \int_T \left(f_1^{(r)}(t) - f_2^{(r)}(t)\right)^2 dt\right]^{1/2}, \qquad [8]$$

where $f_1^{(r)}(t)$ and $f_2^{(r)}(t)$ represent the *r*-derivatives of $f_1(t)$ and $f_2(t)$, respectively. This semi-metric is particularly useful when analysing properties such as the rate of change, acceleration, or higher-order variations in investor behaviour rather than just the original function.

These metrics and semi-metrics are essential to evaluate the continuous differences between investors and their intertemporal choices. By capturing variations in temporal discounting, they provide a foundation for clustering by quantifying similarities and dissimilarities in decision-making patterns.

In addition to analysing individual profiles, these proximity measures can be extended to compare the average temperament profiles. An average profile is defined as the functional mean of the discount functions for a specific investor type (e.g., Artisans, Guardians, Idealists, Rationals). The functional mean $\bar{f}_g(t)$ for a given group $C_g$ is calculated as:

$$\bar{f}_g(t) = \frac{1}{n_g} \sum_{i=1}^{n_g} f_i(t) \qquad [9]$$

where $n_g$ is the number of investor profiles in the group $C_g$. Using the $L_2$-distance or the semimetric of derivatives, we can then measure the distance between two average profiles $\bar{f}_{g_1}(t)$ and $\bar{f}_{g_2}(t)$:

$$d(\bar{f}_{g_1}(t), \bar{f}_{g_2}(t)) = \left[\frac{1}{T} \int_T \left(\bar{f}_{g_1}(t) - \bar{f}_{g_2}(t)\right)^2 dt\right]^{1/2} \qquad [10]$$

This comparison allows us to quantify the dissimilarities between the mean behaviours of different investor temperaments, providing insights into how distinct behavioural types may perceive and evaluate intertemporal choices differently.

To gain deeper insights into investor behaviors, it is crucial not only to measure the proximity between average profiles but also to evaluate the functional variability both within and between temperaments. Functional variability refers to the spread and dispersion of individual curves around the mean profile of a group over time, highlighting how consistent (or divergent) investor profiles are within a specific behavioural type. Mathematically, this variability is quantified using measures such as the functional variance, which captures the degree of fluctuation of individual discount functions around the group mean $\bar{f}_g(t)$.

The within-group variability for a temperament group $C_g$ can be defined as:

$$\text{Var}_g(t) = \frac{1}{n_g} \sum_{i=1}^{n_g} \left[f_i(t) - \bar{f}_g(t)\right]^2 \qquad [11]$$

where $f_i(t)$ represents the discount function of the $i$-th investor within the temperament $C_g$, and $\bar{f}_g(t)$ is the functional mean of the temperament. This measure provides insights into how heterogeneous the discounting patterns are among investors of the same behavioural type. Low variability indicates that most individuals in the group exhibit similar intertemporal preferences, suggesting a solid alignment within that behavioural profile. Conversely, high variability may imply the presence of significant subgroups with distinct attitudes toward time and risk, even within the same temperament. The between-group variability captures differences in decision-making patterns across different temperaments and can be computed as the sum of squared differences between the functional means of each temperament and the global mean:

$$\text{Var}(t) = \sum_{g=1}^{G} \left[\bar{f}_g(t) - \bar{f}(t)\right]^2 \qquad [12]$$

where $\bar{f}_g(t)$ represents the functional mean for the $g$-th temperament group, and $\bar{f}(t)$ is the overall functional mean across all temperaments. This measure allows us to quantify how distinct each temperament group is from the global average in their approach to intertemporal choices. A high between-group variability indicates that the groups are well separated in terms of their temporal discounting behaviours, supporting the hypothesis that different temperaments employ unique strategies when evaluating future outcomes.

Analysis of variability both within and between can be helpful in many ways in the context of intertemporal choices. Indeed, high within-group variability may signal the need for a more granular segmentation of investor types, whereas low between-group variability could indicate overlapping behaviours, suggesting that some groups might be merged. In other words, within-group variability can highlight the presence of sub-clusters that represent distinct subtypes within a broader behavioural category. For example, if Rationals show high variability, it may be due to the existence of distinct "strategic" and "opportunistic" subgroups. Hence, functional variability enables the refinement of behavioural finance theories by testing whether traditional classifications (e.g., Pompian's BITs) accurately capture the diversity of real-world behaviours.

### 4.2 Functional K-means for Clustering Individual Investor Types

Numerous functional clustering techniques, along with various distances and semi-metrics, can be applied depending on the characteristics of the functional data and the study objectives. One widely used approach is the classical functional $k$-means clustering technique, which identifies partitions that minimise cluster variability. Unlike traditional $k$-means clustering, which is applied to finite-dimensional data, the functional variant extends the concept to entire curves, trajectories, or temporal sequences.

Starting from *n* functional observations, the goal is to group these functions into *G* groups, denoted as $C_1, C_2,...,C_G$, to minimise the within-cluster sum of squares. The initial step is to select *G* centroids, denoted as $\psi_1^{(0)}(t), ..., \psi_G^{(0)}(t)$. Each function is then assigned to the cluster whose centroid, at the previous iteration, is the nearest:

$$C_i^{(m)} = \arg \min_{g \in \{1,...,G\}} d^2\left(f_i(t), \psi_g^{(m-1)}(t)\right), \quad m = 1, ..., M \qquad [13]$$

where *M* is the maximum number of steps for the algorithm. Once all functions have been assigned to a cluster, the functional cluster means are updated as follows:

$$\psi_g^{(m)}(t) = \frac{\sum_{f_i(t) \in C_g} f_i(t)}{n_g} \qquad [14]$$

where $n_g$ is the number of functions in the *g*-th cluster, $C_g$.

Hence, the within-cluster sum of squares $WCSS = \sum_{g=1}^{G} \sum_{i \in C_g} d^2(f_i(t), \psi_g(t))$ is to objective function to be minimised; however, the distance function $d^2(f_i(t), \psi_g(t))$ can be chosen based on the specific characteristics of the functional data, such as $L_2$ distance, semi-metric of the *r*-order derivative illustrate in Section 4.1, or even other functional metrics or semimetrics.

The optimal number of clusters in functional *k*-means can be determined using various criteria. One effective approach can be to use an extension of the Elbow method applied to the coefficients of the basis functions (e.g., B-spline coefficients). The classical Elbow method involves plotting WCSS against the number of clusters. The idea is to identify a point where the decrease in WCSS becomes marginal, which indicates the optimal number of groups. Using the B-spline coefficients as features, the same concept can be extended to functional data, making it possible to identify an appropriate number of clusters that balance complexity and homogeneity.

**4.3 Functional *K*-means for Subgroup Identification within Existing Investors' Profiles**

To further explore the internal structure of pre-defined temperaments in intertemporal decision-making, we introduce a functional *k*-means clustering variant conditioned on existing group labels. Given four main classes identified through prior analysis (e.g., behavioural investor types such as Artisans, Guardians, Idealists, and Rationals), the objective is to identify potential subgroups within these larger groups using functional *k*-means. The basic idea is to understand if a functional clustering algorithm can identify heterogeneous behaviours within these identified classes through the typologies identified in the literature.

The proposed conditioned functional *k*-means approach works as follows:

1  The functional observations are treated as separate datasets for each of the four classes.

2. For each class, the functional *k*-means algorithm is applied independently to search for internal patterns or subgroups that may have gone undetected in the primary clustering phase.
3. The number of subgroups $G_j$ is selected for each class $j$, aiming to minimise the WCSS within each group while maximising the separation between subgroups.

Given a set of functional observations $\{f_i(t)\}$ for investors' class $j$, the conditioned functional *k*-means aims to solve the following minimisation problem:

$$\arg\min_{\psi_g^{(j)}(t)} \sum_{g=1}^{G_j} \sum_{i \in C_g^{(j)}} d^2\left(f_i(t), \psi_g^{(j)}(t)\right) \quad [15]$$

where $G_j$ is the number of subgroups in class $j$, $C_g^{(j)}$ represents the set of functional observations in the $g$-th subgroup of class $j$, $\psi_g^{(j)}(t)$ is the centroid function of the $g$-th subgroup, and $d^2\left(f_i(t), \psi_g^{(j)}(t)\right)$ is the distance measure.

The result is a fine-grained partitioning of the four main classes, enabling a deeper understanding of intra-group dynamics and behaviours.

The number of subgroups for each starting investors' class can be selected using various criteria. In section 4.2, an extension of the Elbow method is proposed but a possible alternative can be the method proposed by Maturo and Verde (2023), which extends the silhouette coefficient to the functional context. The average silhouette determines how well each investor profile fits within its cluster and can be calculated for different values of $G$. The optimal number of clusters $G^*$ is the one that maximizes the average silhouette across a range of possible values for $G$. The silhouette for the $i$-th investor profile is computed as follows:

$$S(i) = \frac{b(i) - a(i)}{\max(b(i), a(i))} \quad [16]$$

where:

- $a(i)$ is the average distance of the $i$-th investor profile to all other profiles within the same cluster $C_g$, excluding itself. It is calculated as:

$$a(i) = \frac{1}{n_g - 1} \sum_{j \in C_g} d\left(f_i(t), f_j(t)\right) \quad [17]$$

where $n_g$ is the number of investor profiles in cluster $C_g$.

- $b(i)$ is the minimum distance of the $i$-th investor profile to all profiles belonging to different clusters, defined as:

$$b(i) = \min_l d(f_i(t), f_l(t)) \quad [18]$$

where $l$ represents investor profiles not in $C_g$.

## 5. DATA AND RESULTS
### 5.1 Data

The test was administered using the *vercel.app* platform and disseminated via social platforms (Instagram and Facebook) in February 2023 to all volunteers over 18. Individuals interested in participating consented to process their data and received the link to the questionnaire from the authors. 200 interviews were collected, but only 170 were complete and without writing errors (e.g. if the participants' discount function was greater than 1, they were excluded from the analysis). Sharing through social platforms was taken to diversify the sample as much as possible, varying social and cultural backgrounds. Once the survey was completed, the data was downloaded and converted into an Excel file. The information requested from the user was gender and age. For each section of the questionnaire, an introductory part explained how to answer. The questionnaire consists of three parts: the first two are aimed at attitude for inter-temporal choices, and the last one implements the Keirsey Temperament Sorter (Keirsey, 1998) to calculate the individual's score concerning each temperament.

For the inter-temporal choice tasks, the respondent was informed of the presence of a 20-second timer constantly visible in a box below the question to have a constant view of the passage of time, at the end of which, in the event of a non-response, the following sentence would appear to prompt the individual to enter the figure: "*Time has expired! The time for answering the question has expired. Please answer the question IMMEDIATELY, without further thought*". This experimental dynamic, introduced to convey a sense of haste during the task, is fundamental to the manifestation of the behavioural anomaly to the extent that the instructions emphasise the test's invalidity when the timer expires for the tenth time. The first questionnaire aims to collect the values in Table 1 for the construction of the discount function and consists of two alternating questions in which the second one seeks to distract the respondents from keeping track of the answers given in the previous questions (Ventre and Martino, 2022, Ventre et al., 2022b). The second questionnaire, on the other hand, repeats the questions used in Ventre et al. (2022a, 2023) to calculate the hyperbolic factor (Rohde, 2010) concerning four anomalies of the discounted utility model: the delay effect, the sign effect, the interval effect and the magnitude effect (Read, 2004). Finally, the third questionnaire did not have a 20-second timer, but it did have a 16-minute timer to encourage respondents to complete it without rushing their answers. Table 4 shows the characteristics of the sample.

| N=170 | Rationals | Idealists | Artisans | Guardians | Female | Male | Mean Age |
|---|---|---|---|---|---|---|---|
| | 42.35% | 37.65% | 7.65% | 12.35% | 57.06% | 42.94% | 28 |

**Table 4.** Characteristics of the sample.

### 5.2 Preliminary Descriptive Results

Figure 1 shows the boxplots of the entire sample and highlights a substantial variability between the temperament types. As there are no outliers, the dispersion of the data around the median for each time instant considered is significant and potentially rich in information for the four categories. In general, however, the median represented by the horizontal lines of the box plots shows a decreasing trend over time, in line with empirical predictions of a hyperbolic discount. When observing the medians for $t=2$, $t=4$ and $t=7$, the sample's tendency towards a steep discount is evident, which diminishes as time increases. The box size also decreases as time advances, confirming that the range of the data distribution falls for periods further out in time.

Having made some general remarks on the discount attitude, Figure 2 depicts the box plots for each temperament at each time instant. Through this representation, exciting differences emerge. First, we observe that the variability of the temperaments is very different. The Idealists present the box plots with the most significant variability over time, followed by the Rationals. The variability of the Artisans decreases as time progresses, indicating less dispersion of the data around the median value. Concerning the median, it is interesting to note that the median for the Rationals takes a lower value than other temperaments for $t=2$ and a higher value for $t=4$, reflecting that Independents generally do not believe in following a long-term investment plan. A substantial discount is applied by the Artisan category, whose median takes on a lower value for all time instants considered, an expression of the impulsiveness that characterises Accumulators. Concerning the four temperaments, the presence of outliers is also minimal, with only five values in total being present for the last instants of time.

The interpolation of the discount functions on the discrete set of points in time, shown in Figure 3, highlights the numerous intersections present primarily in the instants of time closer to the present, where emotional drives and behavioural attitudes define a substantial impact on the discount function. Moreover, while the Artisans deviate from the trend of the other categories, the remaining intersect up to $t=60$. These intersections emphasise how much time and the individual's perception of it influences behavioural aspects. To explore the characteristics of the decision-making process in the context of the intertemporal choices of the four temperaments, Figure 4 depicts the graph of uncertainty aversion, and Table 4 shows the medians of the hyperbolic factors. From Figure 4, the higher degree of uncertainty aversion of the Guardian's temperament reflects their financial inertia:

Preservers, for fear of making the wrong decision or taking too many risks, prefer to avoid investing, and this would match the level of uncertainty aversion. Idealists are characterised by the lowest degree of uncertainty aversion, and it could be because Followers are generally disinterested in financial activities. Interestingly, Artisans, on the other hand, are characterised by having a harmful degree of uncertainty aversion for the first period, which expresses an attitude of uncertainty aversion in line with their behavioural characteristic of being risk-prone, in line with the overconfidence of Accumulators, who think they can control the outcome of the investment.

Compared to Table 4, the decrease in impatience highlights the anomalies in the discounted utility model to which the four temperaments are more or less prone. In particular, H(0,6,500,12) refers to the delay effect, H(0,6,50,12) refers to the magnitude effect, H(0,1,50,1) refers to the interval effect, and H(0,6,-500,12) refers to the sign effect (Ventre et al., 2022a). For example, the higher loss tolerance of the Artisans with a hyperbolic factor of 0.67 for H(0,6,-500,12) is evident. By relating the values in Table 4 to each other, it is possible to investigate further how Temperaments respond to anomalies. For example, the interval effect has a strong impact, especially for Artisan, as the hyperbolic factor H(0,1,50,1) increases by about 2.4 compared to H(0,6,50,12) and has minimum effect on the Guardians, who do not tend to change their discounting attitude concerning the interval of choice, since the ratio between H(0,6,50,12) and H(0,1,50,1) is equal to 1: what has been observed confirms the trend of uncertainty aversion for Accumulators, and the attitude of preserving their budget for Preservers. However, the Guardians are very sensitive to the magnitude effect, so the hyperbolic factor H(0,6,500,12) increases by more than ten times when the figure goes from 500 to 50, as shown by the ratio between H(0,6,500,12) and H(0,6,50,12). This result could align with the Preservers' inclination to mental accounting bias whereby, when categorising money, there is a different attitude for smaller amounts as they are not perceived as a loss of their asset, so they take a much more impulsive attitude. Another anomaly that confirms the behavioural description of the Guardians is the sign effect, whereby the hyperbolic factor H(0,6,500,12) increases by about five times when outcomes are negative, i.e. -500 is considered instead of 500, and thus represents losses. Other similar considerations that can be obtained from Table 5 confirm previous studies (Martino et al., 2023; Ventre et al., 2023).

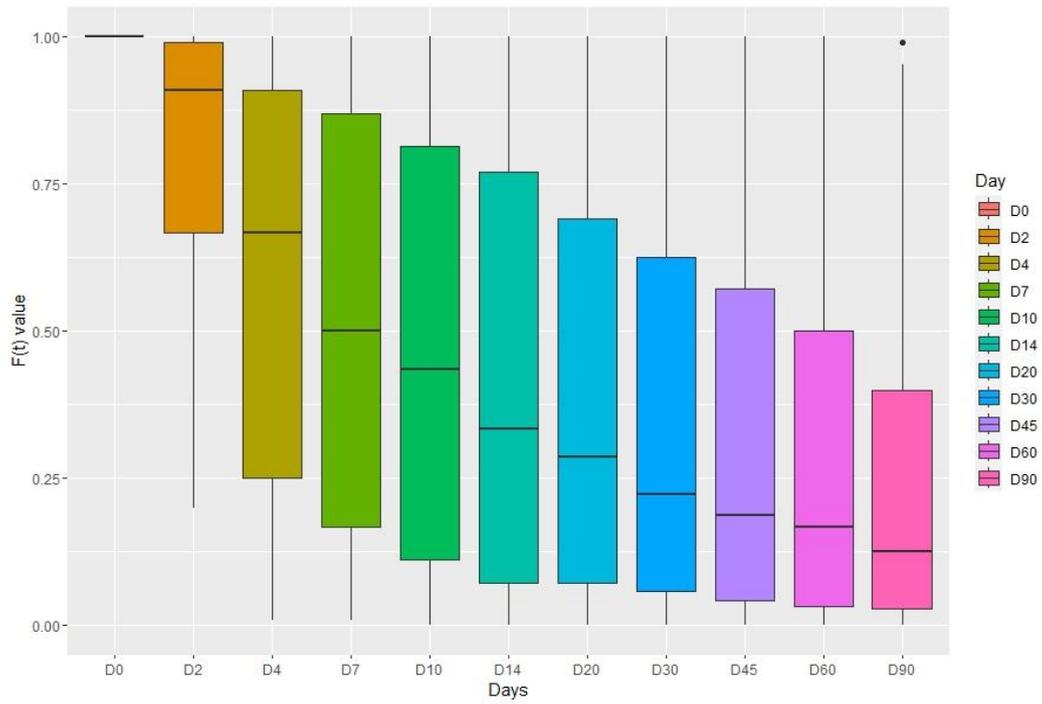

**Figure 1.** Box plots on *f(t)* values at different days.

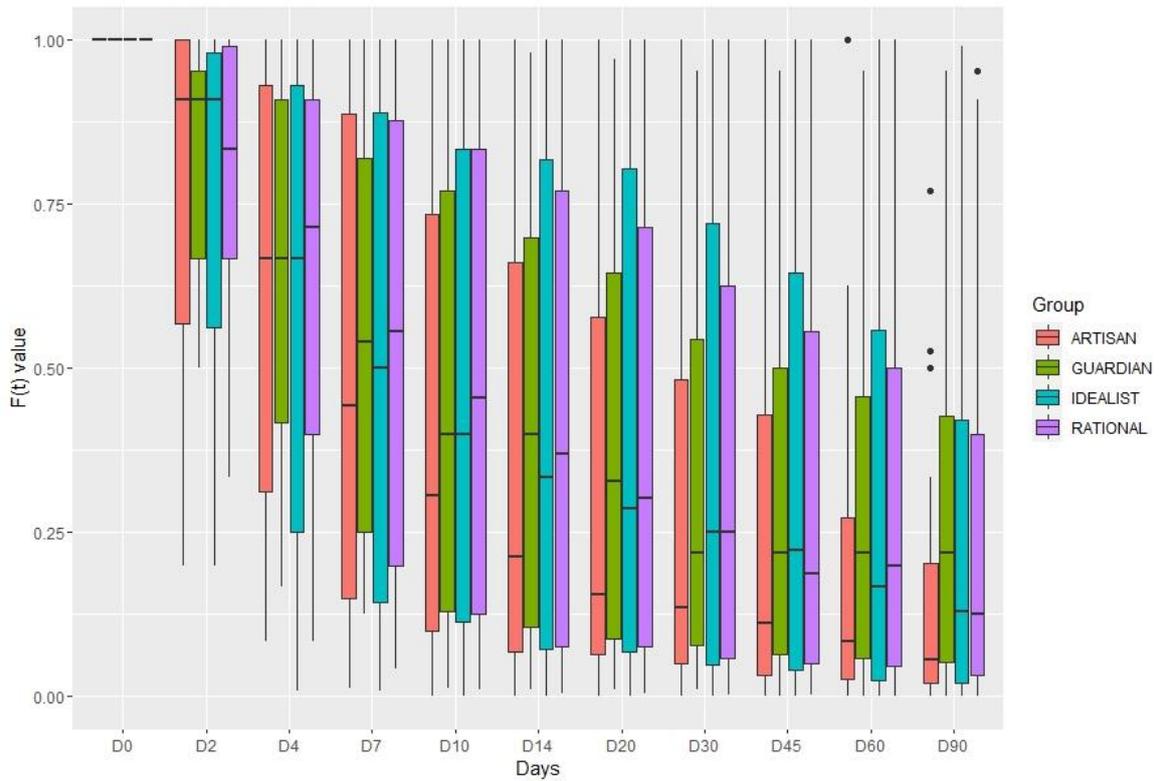

**Figure 2.** Box plots on *f(t)* values at different instants of time for each temperament.

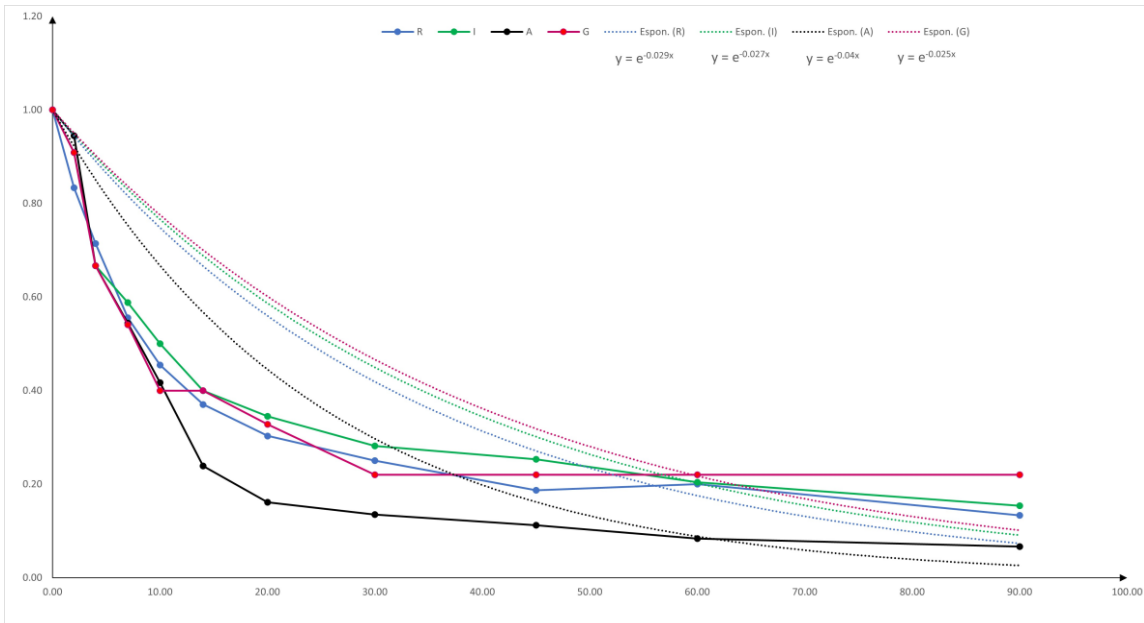

**Figure 3.** The discount function was obtained by interpolating median values of *f(t)* on different days for each temperament. The dotted line represents the exponential approximation of the empirical curves.

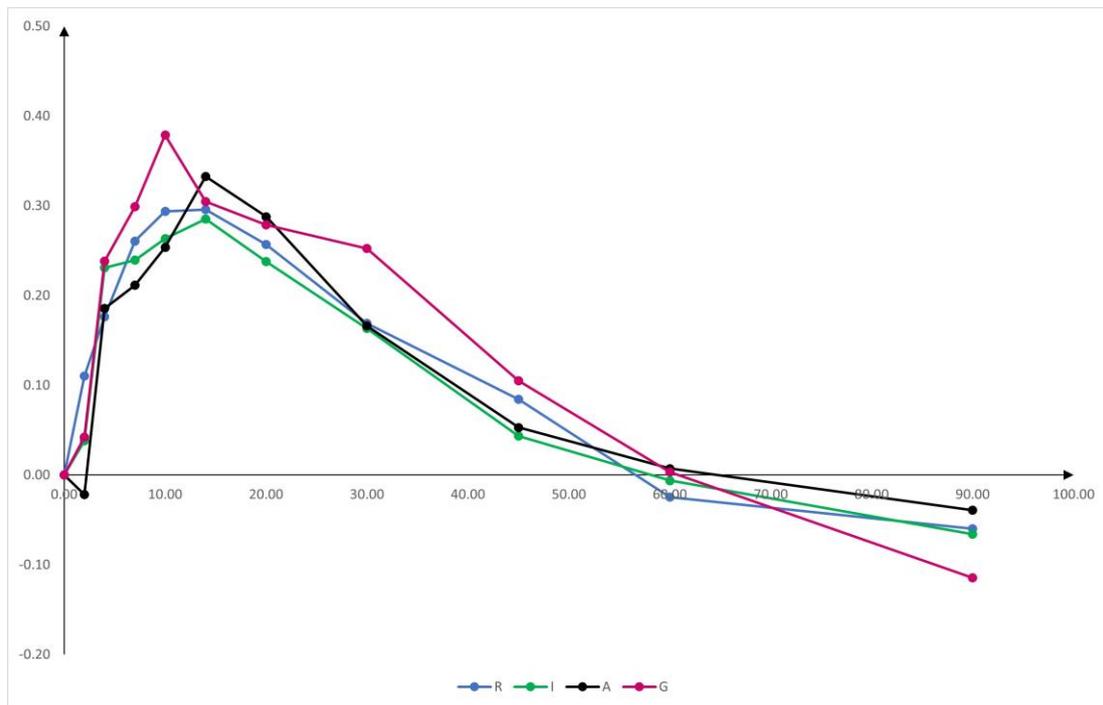

**Figure 4.** The difference between the empirical discount function and the exponential discount function for each temperament.

|  | H(0,6,500,12) | H(0,6,50,12) | H(0,1,50,1) | H(0,6,-500,12) |
|---|---|---|---|---|
| Rational | 0.83 | 1.83 | 4.12 | 1.41 |
| Idealist | 0.50 | 1.11 | 2.69 | 0.83 |
| Artisan | 0.17 | 1.69 | 4.12 | 0.67 |
| Guardian | 0.17 | 1.83 | 1.83 | 0.83 |

Table 5. Medians of the hyperbolic factors.

### 5.3 Functional Data Analysis Results

FDA allows for even more detailed representations, defined by moving from discrete to continuous time intervals, representing intertemporal preferences as functions rather than point data. FDA enables the evaluation of significant additional sources of patterns and variations in the data (Ferraty and Vieu, 2006; Maturo and Porreca, 2022); indeed, often crucial information resides within the first and second derivatives rather than solely within the raw data. Furthermore, FDA offers the theoretical capability to observe the phenomenon at a much finer grid and, in theory, to follow *f(t)* at any fixed moment *t* (see, e.g., Ramsay and Silverman 2005; Ferraty and Vieu, 2006; Maturo et al. 2019a, 2019b; Maturo and Verde 2022).

Figure 5 accurately depicts the discount functions for each temperament, increasing the quality of the analysis. Many more intersections emerge in the first period, and the behaviour described in the continuum emphasises the detachment of the Artisans and the similarity in attitude between the Rationals and the Idealists. To investigate how time influences the decision maker's perspective, with FDA, it was possible to obtain the first derivative of the discount function for each temperament, shown in Figure 6. The analysis of the first derivative shows a symmetry between the Artisan and the Guardians, which is evident up to about 20 days. Furthermore, the fact that the first derivative functional mean of the Guardians varies less than that of the other temperaments in the first period could align with the above result about the interval effect in Table 4, whereby the discount function in periods close to the present decreases less. Also, Figure 5 shows possible inflexion points for the discount functions. To investigate this, Figure 7 represents the second derivative of the discount function. The Guardian category shows variations in the discount function over a more extended period, as the last maximum point is at $t = 10$. Furthermore, while the second derivative of the discount functions increases from $t = 0$ to $t = 2$ for all temperaments, that associated with Guardians decreases over this interval and cancels out over time several times, indicating a continuous change in intertemporal preference patterns.

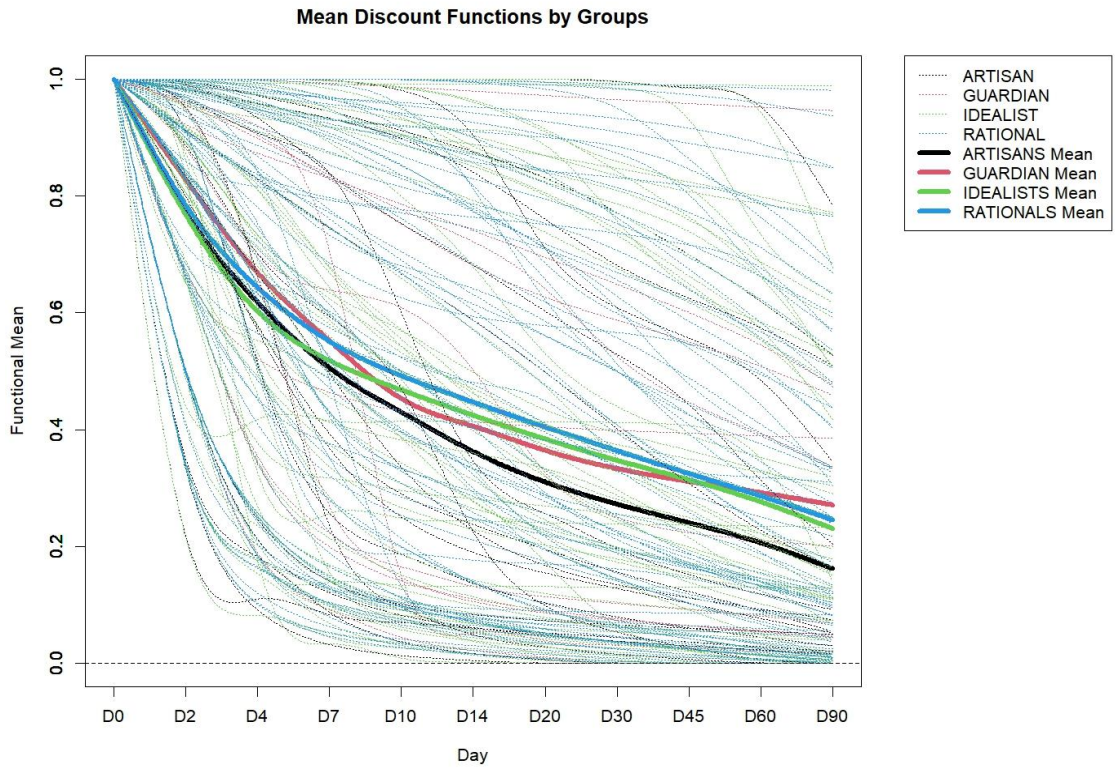

**Figure 5.** Discount function for each temperament.

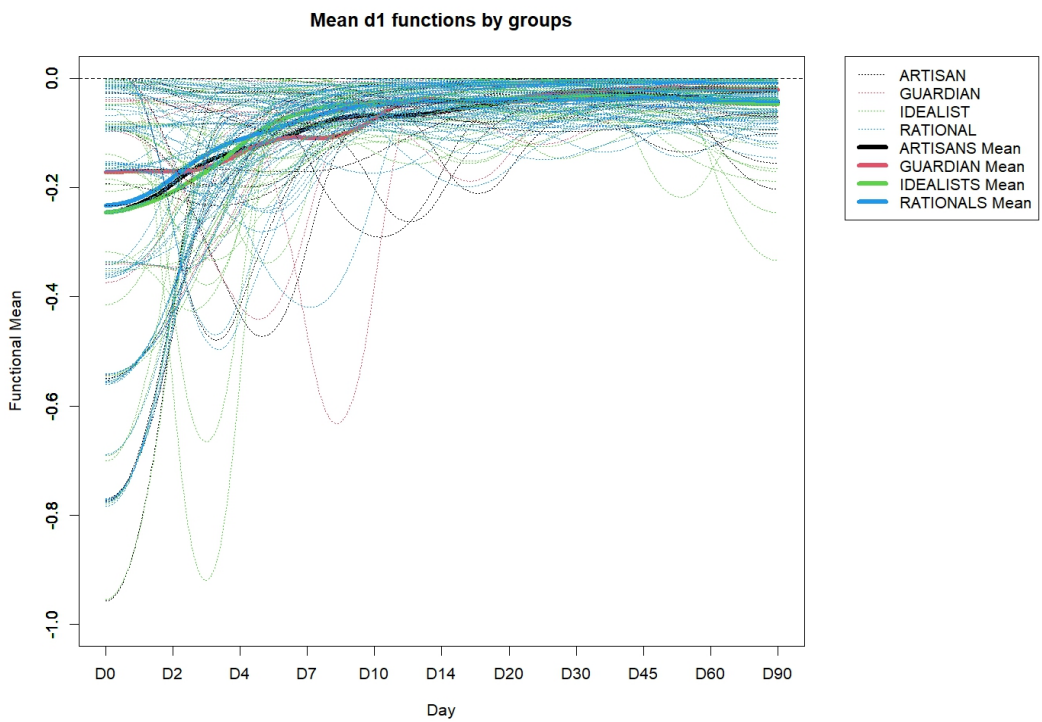

**Figure 6.** First derivatives of the discount functions for each temperament.

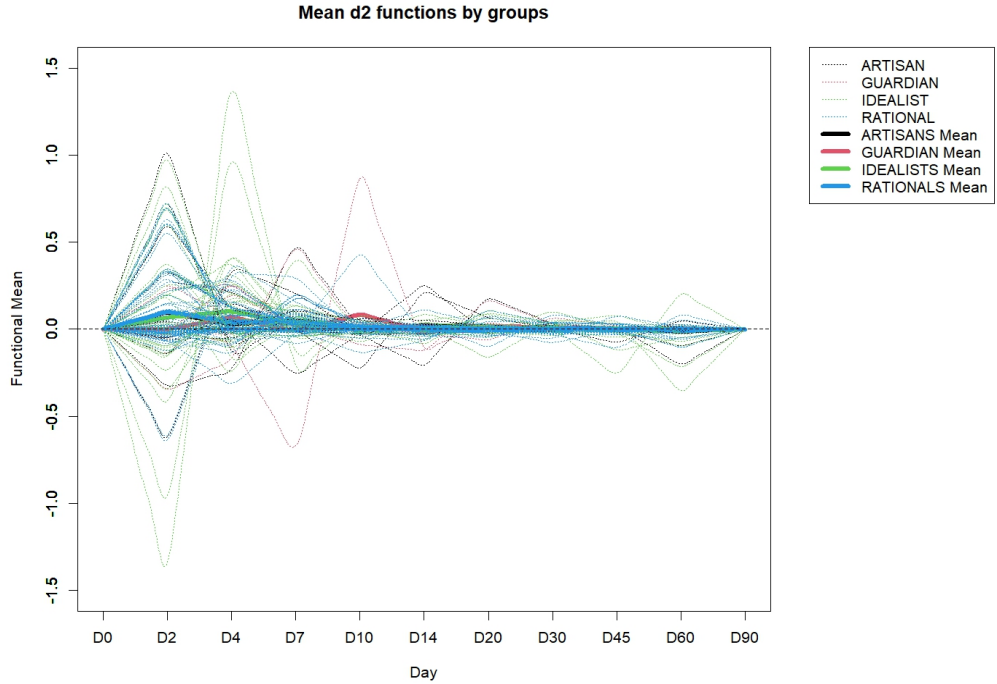

**Figure 7.** Second derivatives of the discount functions for each temperament.

### 5.4 FUNCTIONAL VARIABILITY AND CLUSTERING RESULTS

The preliminary analysis is a comprehensive exploration of within-group variability, providing a deep understanding of the heterogeneity of profiles within each temperament. High functional variability may indicate the presence of subgroups within a given temperament, revealing underlying behavioural distinctions that are not immediately apparent. These findings have intriguing implications for understanding the complex nature of intertemporal choice profiles and their variability. Figure 8 displays the within-group functional variance for each temperament over time. The curves illustrate the internal variability of behavioural profiles for the four identified groups: Artisan, Guardian, Idealist, and Rational. The functional variances follow an increasing trend in the early days, peaking around day 14-30, and then slightly decreasing towards day 90. This pattern suggests that the intertemporal choice profiles of each temperament exhibit higher dispersion during the middle phase of the considered period and tend to converge in the initial and final stages. The variance for each group reveals distinctive characteristics, each with its implications. The Artisan (black) group displays the highest variance during the first phase, indicating greater heterogeneity in intertemporal discounting preferences compared to the other groups. The Guardian (red) curve shows a slower initial growth with an intermediate peak, suggesting greater stability in temporal preferences during the early days.

Figures 9 and 10 illustrate the functional within-group variance curves for each temperament group, considering the first and second derivatives of the discount functions. Figure 9 shows the within-group variance for the first derivative, which provides insights into the velocity of profile changes over time. Initially, all temperaments exhibit a relatively high within-group variability, especially the Artisan. This variance decreased significantly over time, indicating the convergence in profile behaviours' rate of change. Figure 10 depicts the within-group variance for the second derivatives, corresponding to the acceleration of profile changes. The Idealist group has the highest initial variability in this figure, suggesting a less homogeneous response pattern than other temperaments. The variance of acceleration for all groups gradually diminishes over time, reflecting a stabilisation of individual trajectories.

These visualisations support the preliminary assessment of each temperament's functional characteristics, providing a nuanced view of how groups differ in their baseline functions and evolution of their change rates and acceleration patterns over time.

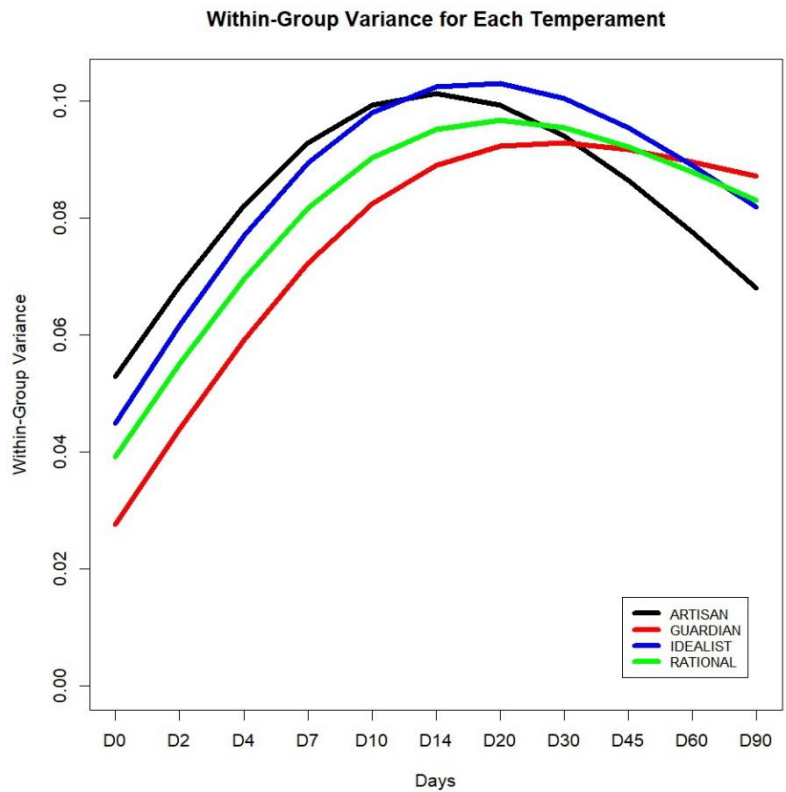

**Figure 8.** Within-group functional variance for each temperament, highlighting internal heterogeneity and potential subgroups over time

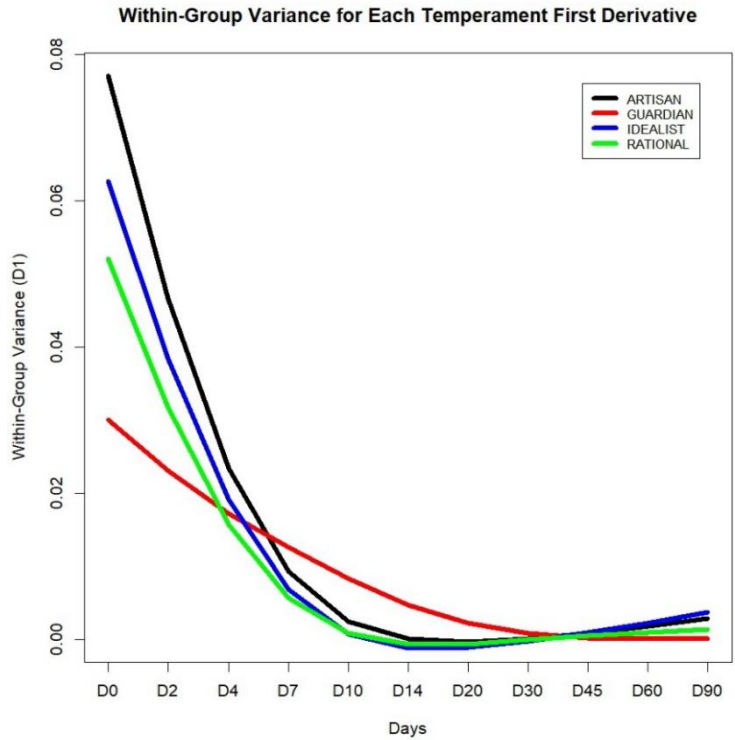

**Figure 9.** Within-group functional variance for each temperament's first derivative, highlighting internal heterogeneity and potential subgroups over time for profile velocity.

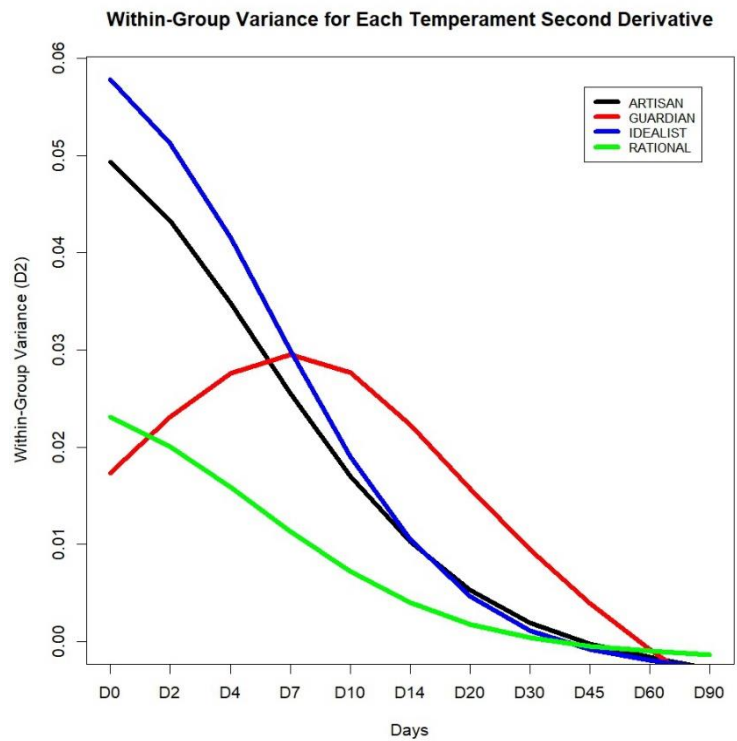

**Figure 10.** Within-group functional variance for each temperament's second derivative, highlighting internal heterogeneity and potential subgroups over time for profile acceleration.

After performing these preliminary evaluations, we can investigate whether subpatterns effectively exist within each temperament using clustering techniques. The idea is to verify if the within-group variability translates into distinct subgroups. By applying functional clustering methods, we aim to identify potential underlying profiles that might otherwise be hidden within the broader temperamental categories.

We can proceed by using the metric and the two semimetrics (considering the first two derivatives) introduced in Equations 7 and 8. However, with four subgroups, this would result in twelve different clustering outcomes. While this is interesting, it would lead to excessive results and figures. For this reason, we limit our analysis to using the metric defined in Equation 7 and proceed with identifying potential subgroups for the four temperaments using the original functions. We removed functional outliers from the subgroups because the $k$-means algorithm is sensitive to anomalous values, as it relies on the mean curve of each group. To identify outliers, we used the Modified Band Depth (MBD) method, which measures the depth of each curve relative to the others in the group. Outliers were defined as curves falling outside the central 30% region, indicating a significant deviation from the typical behaviour.

Figure 10 displays the results of the Elbow method applied to each of the four original temperamental classes: Artisan, Guardian, Idealist, and Rational. This method aims to identify the optimal number of subclusters within each temperament by examining the Within-Cluster Sum of Squares (WCSS) as a function of the number of subclusters. The WCSS measures the compactness of clusters, and a significant drop, or "elbow," indicates the point where adding more clusters does not substantially improve the clustering results. The vertical red line on each plot indicates each temperament type's chosen number of subclusters. The Elbow plots show that the original classes are quite internally heterogeneous. For instance, the Artisan and Idealist groups show significant drops in WCSS up to four subclusters, indicating considerable variability among individual profiles within these temperamental types. Similarly, the Guardian and Rational classes show a marked heterogeneity, requiring three and four subclusters to capture the internal variability adequately. These results suggest that the original classes encompass various subgroups of investors, each with distinct temporal decision-making behaviours. Given the high internal variability, interpreting these groups' global means (centroids) may be more straightforward than analysing the individual subclusters' centroids. This complexity highlights that different original temperaments contain subgroups of investors with markedly distinct patterns in their intertemporal choices.

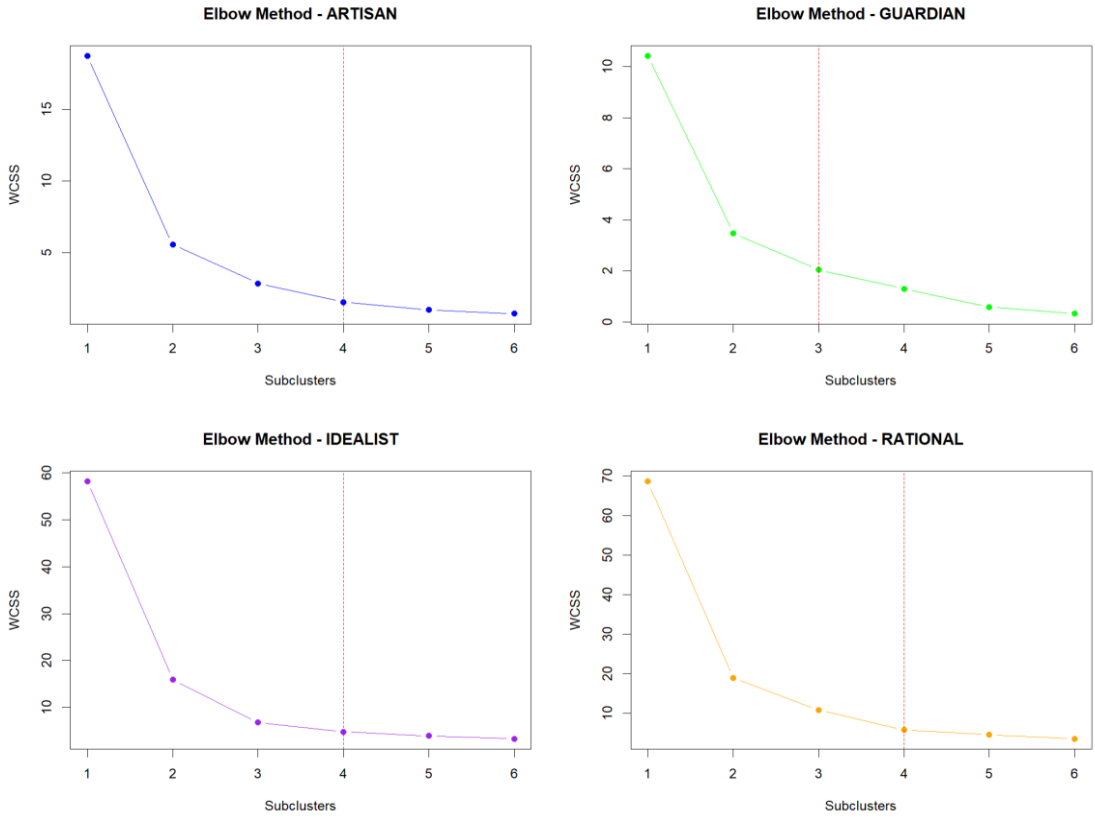

**Figure 10.** Selecting the number of sub-temperaments using an extension of the Elbow method.

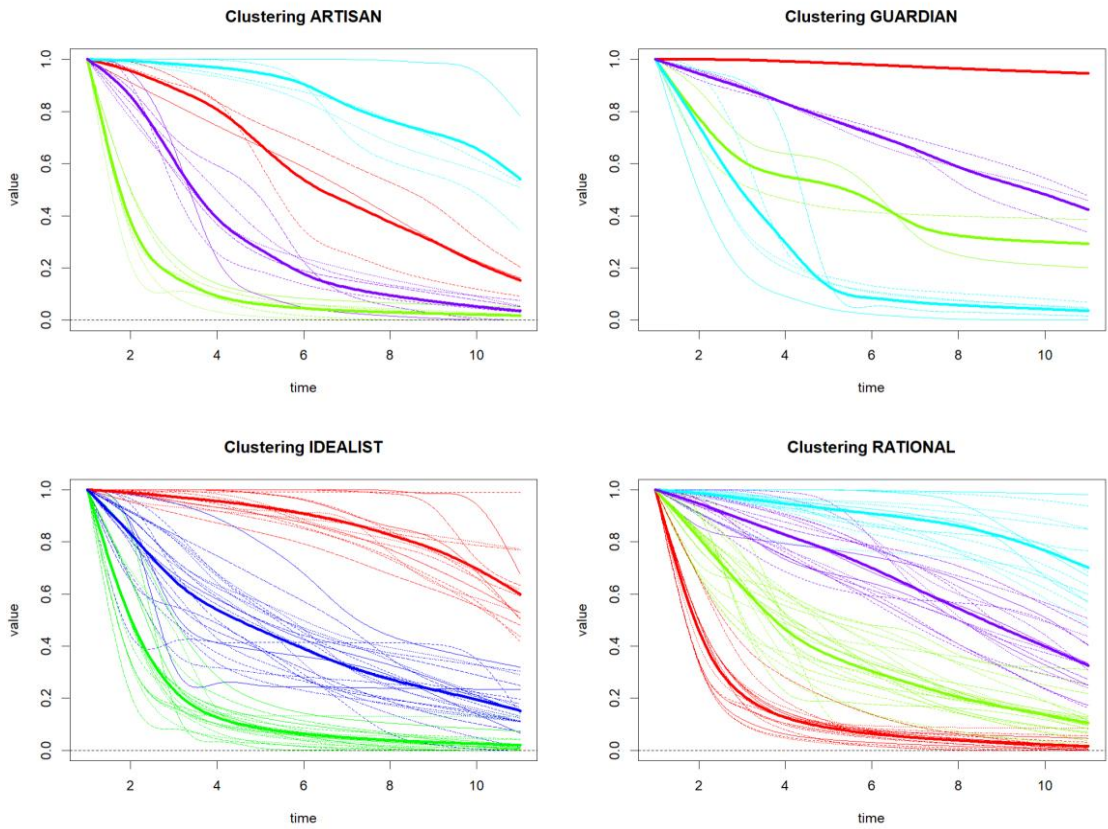

**Figure 11.** Sub-temperaments functional centroids using conditional functional k-means.

In light of the findings from the conditional functional k-means clustering, based on the known temperamental classes, an additional analysis of interest involves understanding which types of investors share similar intertemporal choice patterns independent of the temperamental categories already established in the literature. The goal is to explore whether new patterns or types of investors can be identified beyond the four predefined temperaments. This analysis could reveal new subgroups with shared behavioural tendencies that have not yet been categorised.

To this end, we implement a functional *k*-means clustering method without considering any a priori knowledge of the temperamental classes, allowing the data itself to define the investor profiles. This unsupervised approach aims to uncover latent structures in intertemporal decision-making behaviours, potentially leading to new insights into how different investor profiles emerge based solely on their temporal choices. Identifying these profiles may help refine the understanding of investor behaviour and guide more personalised financial strategies that better align with individual preferences and decision-making tendencies.

First, we select the number of clusters. By applying the Elbow method to the global functional dataset, we can determine the optimal number of clusters for grouping the entire investor population based on their intertemporal choice patterns independently of predefined temperamental classes.

Figure 12 shows the Elbow Method applied to all combined curves without considering the prior classifications. As depicted in the graph, the optimal number of clusters is four, marked by the red dashed line. This means four distinct behavioural groups emerge when combining all functional curves without considering the original classifications. This result suggests that, despite the original predefined classes, the optimal functional *k*-means clustering identifies four cohesive patterns of temporal decision-making.

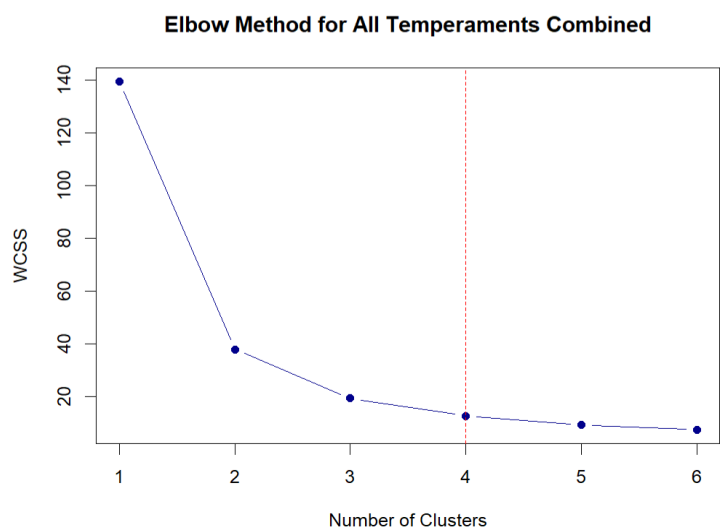

**Figure 12.** Selecting the number of investor types using an extension of the Elbow method**.**

Figure 13 presents the outcome of a functional *k*-means clustering analysis applied to investor discount curves without considering predefined temperament categories. This approach reveals four distinct patterns, emphasising the complexity and heterogeneity in intertemporal decision-making that are not captured by traditional temperament-based groupings. The four classes identified are clearly different from the four temperaments analyzed previously.

The green cluster, characterised by steeply declining discount curves, represents investors with a strong preference for immediate gains and a tendency to undervalue future outcomes heavily. This behaviour suggests impulsivity, where short-term rewards are highly valued over long-term benefits. In contrast, the cyan cluster displays a more gradual decrease in value, indicating a more balanced approach. Although these investors still slightly prefer short-term returns, they are more open to considering long-term options.

The purple cluster has a flat discount function, showing that investors in this group maintain a stable valuation for future rewards, making them more patient and long-term oriented. Such behaviour is typical of rational decision-makers who do not devalue future outcomes excessively. Finally, the red cluster depicts a unique dynamic, where discount rates initially drop sharply and stabilize. This pattern suggests that these investors initially prefer short-term gains but can shift their focus toward long-term rewards if immediate returns are not sufficiently attractive.

The new clusters highlight the diversity within investor preferences and suggest that the original temperaments may oversimplify the range of behaviours. By using functional *k*-means clustering without a priori knowledge, we capture a more detailed view of investor profiles, uncovering patterns that would remain hidden within traditional classifications. This refined classification contributes to a better understanding of time preferences and can inform more tailored investment strategies. Consequently, the resulting insights challenge the adequacy of traditional categories and underscore the need for a deeper exploration of temporal dynamics in investment behaviour.

## 6. DISCUSSION AND CONCLUSIONS

The results of this study provide new insights into the heterogeneity of behavioural investor types (BITs) by using Functional Data Analysis (FDA) and clustering techniques. The main contribution of the paper lies in its ability to capture the underlying complexity of temporal preferences, which are often oversimplified in traditional behavioural finance models. Our findings reveal that the application of FDA to the study of intertemporal choices highlights distinct behaviours within each of the four classic BIT categories, such as Accumulators, Preservers, Followers, and Independents.

This approach allowed us to identify not only broader patterns of temporal discounting but also nuanced differences within these groups.

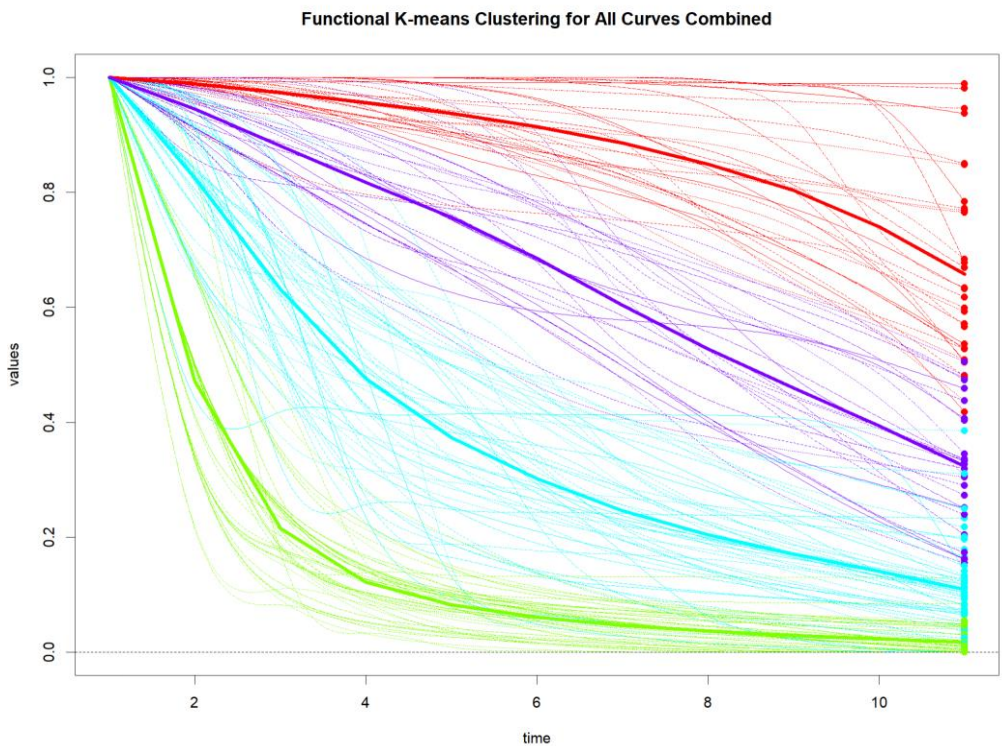

**Figure 13.** Investors' behaviours patterns without considering pre-defined temperaments using unconditional functional *k*-means.

One of the key novelties of our work is the introduction of functional clustering to uncover hidden subgroups within the main behavioural profiles. Traditional models, such as those proposed by Pompian, rely on a fixed number of categories, often failing to account for the variability in investors' responses to risk and uncertainty. By analysing the functional representations of discount functions, we observed that significant heterogeneity exists even among investors classified within the same behavioural category. For instance, the subgroups identified within the Guardian type demonstrated a broad range of risk attitudes, suggesting that some investors might have been incorrectly classified under a single type in previous models. Similarly, the Artisan group, traditionally associated with impulsive behaviours, exhibited distinct temporal profiles that reveal varying degrees of overconfidence and short-term orientation.

These results confirm the hypothesis that classical BITs are too rigid to encompass the complexity of real-world investor behaviour. The introduction of functional clustering thus represents a methodological advancement, providing a more flexible and data-driven approach to investor profiling. In particular, the observed variability in the first and second derivatives of discount functions for each temperament highlights the dynamic nature of decision-making processes and the interplay between temporal preferences and emotional drives. This richer understanding of behavioural patterns opens up new avenues for refining investor classifications, suggesting that a more fluid model, capable of adapting to changes in temporal preferences, is needed. Moreover, the comparison between the empirical discount functions and their first and second derivatives provides evidence that temporal preferences are not static but evolve according to specific behavioural traits and situational contexts.

The implications of these findings are twofold. First, from a theoretical standpoint, they suggest that investor classification models must move beyond static categorizations and incorporate more dynamic methods such as FDA to account for the evolution of preferences over time. Second, from a practical perspective, financial advisors and investment professionals can use these insights to better understand their clients' temporal discounting behaviour, allowing for more tailored investment strategies that reflect the diversity of attitudes within each behavioural type.

In conclusion, the present study successfully demonstrates the potential of combining intertemporal choice theory with FDA and clustering techniques to reveal hidden structures within investor types. This approach challenges the adequacy of traditional BIT classifications and underscores the need for more adaptive models that can capture the complexity and variability of real-world decision-making. Future research should focus on applying these methods to larger and more diverse samples to validate the generalizability of our findings and explore the role of other behavioural dimensions in shaping temporal preferences.

**Funding and/or Conflicts of Interest/Competing Interests**

The authors confirm they received no support from any organisation for the submitted work. They also declare no affiliations or involvement with any organisation or entity with a financial or non-financial interest in the subject matter of this manuscript.

**Use of Generative AI in Scientific Writing**

**Data Availability Statement**

Data can be provided upon request.

# REFERENCES

Cervellati, E. M. (2017). The importance of investors' personality in financial education. *Challenges in ensuring financial competencies*, 77-87.

Cervellati, E. M. (2018). Behavioral Re-Evolution: How Behavioral Economics has Evolved and is Evolving. In *The Behavioural Finance Revolution: A New Approach to Financial Policies and Regulations*. Edward Elgar Publishing Limited, Cheltenham (UK).

Cruz Rambaud, S., & Muñoz Torrecillas, M. J. (2016). Measuring impatience in intertemporal choice. PLoS One, 11(2), e0149256.

Ferraty, F., & Vieu, P. (2006). Nonparametric Functional Data Analysis. Springer. https://doi.org/10.1007/0-387-36620-2.

Frederick, S., Loewenstein, G., & O'donoghue, T. (2002). Time discounting and time preference: A critical review. *Journal of Economic Literature*, *40*(2), 351-401.

Geng, X., Li, M., Zhang, F., Li, W., & Liu, D. (2022). Incremental theory of personality attenuates the effect of environmental uncertainty on intertemporal choices. *Journal of Pacific Rim Psychology*, *16*, 18344909221139325.

Green, L., Myerson, J., & McFadden, E. (1997). Rate of temporal discounting decreases with amount of reward. *Memory & Cognition*, *25*, 715-723.

Levorin, G. (2021). An Empirical Study on Objective and Subjective Financial Risk Tolerance: Moderated Mediation of Financial Literacy and Personality Types.

Loewenstein, G., & Prelec, D. (1992). Anomalies in intertemporal choice: Evidence and an interpretation. *The Quarterly Journal of Economics*, *107*(2), 573-597.

Martino, R., & Ventre, V. (2023). An Analytic Network Process to Support Financial Decision-Making in the Context of Behavioural Finance. *Mathematics*, *11*(18), 3994.

Martino, R., Ventre, V., & di Tollo, G. (2023). Analytic Hierarchy Process for classes of economic behavior in the context of intertemporal choices. *Ratio Mathematica*, *47*.

Joo, B. A., & Durri, K. (2015). Comprehensive review of literature on behavioural finance. *Indian Journal of Commerce and Management Studies*, *6*(2), 11-19.

Kahneman, D. & Tversky, A. (1979). *Prospect theory: an analysis of decision under risk*, *47*.

Kaplan, B. A., Lemley, S. M., Reed, D. D., & Jarmolowicz, D. P. (2014). 21-and 27-item Monetary Choice Questionnaire automated scorers. [software]. Center for Applied Neuroeconomics, University of Kansas.

Kaplan, B. A., Amlung, M., Reed, D. D., Jarmolowicz, D. P., McKerchar, T. L., & Lemley, S. M. (2016). Automating scoring of delay discounting for the 21-and 27-item monetary choice questionnaires. *The Behavior Analyst*, *39*, 293-304.

Keidel, K., Rramani, Q., Weber, B., Murawski, C., & Ettinger, U. (2021). Individual differences in intertemporal choice. *Frontiers in Psychology*, *12*, 991.

Keirsey, D., & Bates, M. (1984). *Please understand me: Character & temperament types*. Prometheus Nemesis Book.

Keirsey, D. (1998). Please understand me II: Temperament, character, intelligence. *(No Title)*.

Kirby, K. N., Petry, N. M., & Bickel, W. K. (1999). Heroin addicts have higher discount rates for delayed rewards than non-drug-using controls. *Journal of Experimental Psychology: General*, *128*(1), 78.

Martino, R., Porreca, A., Ventre, V. et al. Exploring intertemporal decision-making dynamics through functional data analysis: investigating variations in different discount function's dimensions. Qual Quant (2024). https://doi.org/10.1007/s11135-024-01869-y.

Maturo, F., Balzanella, A., & Di Battista, T. (2019). Building Statistical Indicators of Equitable and Sustainable Well-Being in a Functional Framework. *Social Indicators Research*, 146(3), 449–471. https://doi.org/10.1007/s11205-019-02137-5.

Maturo, F., Ferguson, J., Di Battista, T., & Ventre, V. (2019). A fuzzy functional k-means approach for monitoring Italian regions according to health evolution over time. *Soft Computing*, 24(18), 13741–13755. https://doi.org/10.1007/s00500-019-04505-2.

Maturo, F., & Porreca, A. (2022). Augmented Functional Analysis of Variance (A-fANOVA): Theory and application to Google Trends for detecting differences in abortion drugs queries. Big Data Research, 30, 100354. https://doi.org/10.1016/j.bdr.2022.100354.

Maturo, F., & Verde, R. (2022). Supervised classification of curves via a combined use of functional data analysis and tree-based methods. *Computational Statistics*, 38(1), 419–459. https://doi.org/10.1007/s00180-022-01236-1.

Maturo, F., & Verde, R. (2024). Combining unsupervised and supervised learning techniques for enhancing the performance of functional data classifiers. *Computational Statistics*, 39, 239–270. https://doi.org/10.1007/s00180-022-01259-8

McKenna, J., Hyllegard, K., & Linder, R. (2003). Linking psychological type to financial decision-making. *Journal of Financial Counseling and Planning*, *14*(1).

Myers, I. B., McCaulley, M. H., Quenk, N. L., & Hammer, A. L. (1998). *MBTI manual: A guide to the development and use of the Myers-Briggs Type Indicator*. Consulting Psychologists Press.

Rao, A. S., & Lakkol, S. G. (2022). A review on personality models and investment decisions. *Journal of Behavioral and Experimental Finance*, *35*, 100691.

Pan, C. H., & Statman, M. (2013). Investor personality in investor questionnaires. *Journal of Investment Consulting*, *14*(1), 48-56.

Parsaeemehr, M., Rezeai, F., & Sedera, D. (2013). Personality type of investors and perception of financial information to make decisions. *Asian Economic and Financial Review*, *3*(3), 283.

Pompian, M. M. (2012). *Behavioral finance and investor types: managing behavior to make better investment decisions*. John Wiley & Sons.

Pompian, M. (2016). *Risk profiling through a behavioral finance lens*. CFA Institute Research Foundation.

Pompian, M. M. (2017). Risk tolerance and behavioral finance. *Investments and Wealth monitor, Boston*, *20*(31), 34-45.

Ramsay, J.O., & Silverman, B. (2005) Functional Data Analysis, 2nd ed., Springer International Publishing: New York.

Read, D. (2004). Intertemporal choice. In D. J. Koehler & N. Harvey (Eds.), *Blackwell handbook of judgment and decision making* (pp. 424–443). Blackwell Publishing.

Robbins, M., & Ross, C. (2020). Keirsey temperament sorter. *Encyclopedia of Personality and Individual Differences*, 2518-2521.

Rohde, K. I. (2010). The hyperbolic factor: A measure of time inconsistency. *Journal of Risk and Uncertainty*, *41*, 125-140.

Samuelson, P. A. (1937). A note on measurement of utility. *The Review of Economic Studies*, 4(2), 155-161.

Samuelson, P. A. (1952). Probability, utility, and the independence axiom. *Econometrica: Journal of the Econometric Society*, 670-678.

Simon, H. A. (1990). Bounded rationality. *Utility and probability*, 15-18.

Statman, M., & Wood, V. (2004). Investment temperament. *Journal of Investment Consulting*, *7*(1), 55-66.

Takahashi, T., Oono, H., & Radford, M. H. (2008). Psychophysics of time perception and intertemporal choice models. *Physica A: Statistical Mechanics and its Applications*, 387(8-9), 2066-2074.

Tversky, A., & Kahneman, D. (1989). Rational choice and the framing of decisions. In *Multiple criteria decision making and risk analysis using microcomputers* (pp. 81-126). Berlin, Heidelberg: Springer Berlin Heidelberg.

Ventre, V., & Martino, R. (2022). Quantification of Aversion to Uncertainty in Intertemporal Choice through Subjective Perception of Time. *Mathematics*, 10(22), 4315.

Ventre, V., Cruz Rambaud, S., Martino, R., & Maturo, F. (2022a). An analysis of intertemporal inconsistency through the hyperbolic factor. *Quality & Quantity*, 1-28.

Ventre, V., Martino, R., & Maturo, F. (2022b). Subjective perception of time and decision inconsistency in interval effect. *Quality & Quantity*, 1-26.

Ventre, V., Martino, R., Cruz Rambaud, S., Maturo, F., & Porreca, A. (2024). An original approach to anomalies in intertemporal choices through functional data analysis: Theory and application for the study of Hikikomori syndrome. Socio-Economic Planning Sciences, 92, 101840. https://doi.org/10.1016/j.seps.2024.101840.

Ventre, V., Cruz Rambaud, S., Martino, R., & Maturo, F. (2023). A behavioral approach to inconsistencies in intertemporal choices with the Analytic Hierarchy Process methodology. *Annals of Finance*, *19*(2), 233-264.

Wheeler, P., Jessup, C., & Martinez, M. (2002). The Keirsey Temperament Sorter: Investigating the impact of personality traits in accounting. In *Advances in accounting behavioral research* (pp. 247-277). Emerald Group Publishing Limited.

Yilmaz, M., & O'Connor, R. V. (2015). Understanding personality differences in software organisations using Keirsey temperament sorter. *IET Software*, *9*(5), 129-134.